\documentclass[aps,prx,twocolumn,superscriptaddress,showpacs,floatfix]{revtex4-2}
\usepackage{comment}
\usepackage{physics}
\usepackage{amsmath,amsthm,amssymb,amsfonts, color, comment, graphicx}
\usepackage{xcolor}
\usepackage{float}
\usepackage{graphicx}
\usepackage{dcolumn}
\usepackage{bm}
\UseRawInputEncoding
\usepackage{hyperref}

\usepackage[normalem]{ulem}

\newcommand{\bea}{\begin{eqnarray}}
\newcommand{\eea}{\end{eqnarray}}

\newcommand{\br}{\mathbf{r}}

\newcommand{\be}{\begin{equation}}
\newcommand{\ee}{\end{equation}}
\newcommand{\bk}{{{\bf{k}}}}

\newcommand{\beal}{\begin{align}}
\newcommand{\eeal}{\end{align}}

\newcommand{\dg}{{\dagger}}
\newcommand{\pdg}{{\phantom\dagger}}

\newcommand{\btjstrw}{\mathrel{{\rotatebox[origin=c]{90}
{$\bowtie$}}\kern-0.18em\raisebox{-.95ex}{$\bullet$}
\kern-0.5em\raisebox{.97ex}{$\bullet$}
\kern-1.12em\raisebox{.97ex}{$\bullet$}
\kern-0.52em\raisebox{-.95ex}{$\bullet$}}}

\newcommand{\btjnbrR}{{\mathrel{\rotatebox[origin=c]{90}
{$\bowtie$}}\kern-0.22em\raisebox{.9ex}{$\bullet$}
\kern-1.em\raisebox{-.8ex}{$\bullet$}}}
\newcommand{\btjnbrL}{{\mathrel{\rotatebox[origin=c]{90}
{$\bowtie$}}\kern-0.22em\raisebox{-.8ex}{$\bullet$}
\kern-1.em\raisebox{+.9ex}{$\bullet$}}}

\def\a{\alpha}

\def\d{\delta}
\def\e{\epsilon}

\def\m{\mu}
\def\n{\nu}

\def\s{\sigma}
\def\t{\tau}

\def\w{\omega}

\newcommand{\llangle}[1][]{\savebox{\@brx}{\(\m@th{#1\langle}\)}%
  \mathopen{\copy\@brx\kern-0.5\wd\@brx\usebox{\@brx}}}
\newcommand{\rrangle}[1][]{\savebox{\@brx}{\(\m@th{#1\rangle}\)}%
  \mathclose{\copy\@brx\kern-0.5\wd\@brx\usebox{\@brx}}}

\begin{document}


\title{Nematic order in topological SYK models}
\author{Andrew Hardy}
\author{Anjishnu Bose}%
\author{Arun Paramekanti}
\affiliation{Department of Physics, University of Toronto, 60 St. George Street, Toronto, ON, M5S 1A7 Canada}

\date{\today}

\begin{abstract}
We study a class of multi-orbital models based on those proposed by Venderbos, Hu, and Kane which exhibit an interplay of topology, interactions, and fermion incoherence.
In the non-interacting limit, these models exhibit trivial and Chern insulator phases with Chern number $C \geq 1$ bands as determined by the relative angular momentum of the participating orbitals. 
These quantum anomalous Hall insulator phases are separated by topological transitions
protected by crystalline rotation symmetry, featuring Dirac or quadratic band-touching points.
Here we study the impact of Sachdev-Ye-Kitaev (SYK) type interactions on these lattice 
models. 
Given the random interactions, these models display `average symmetries' upon disorder averaging, including a charge conjugation symmetry, so they behave as
interacting models in topological class $\mathbf{D}$ enriched by crystalline rotation symmetry.
The phase diagram of this model features a non-Fermi liquid at high temperature and an `exciton condensate'
with nematic transport at low temperature.
We present results from the free-energy, spectral functions, and the anomalous Hall resistivity as a function of temperature and tuning parameters.
Our results  are broadly relevant to correlated topological matter in multiorbital systems, and may also be viewed, with a suitable particle hole transformation, as an exploration of strong interaction effects on mean-field topological superconductors.
\end{abstract}

\maketitle

\section{\label{sec:level1}Introduction}

Advances in the fields of topological phases and strongly-correlated electron systems have provided exceptions to the two paradigms proposed 
by Landau to characterize condensed matter systems.  Landau theory for symmetry-breaking phase transitions had to be modified 
to include topological phase transitions which are not characterized by a local order parameter. These include classical phase transitions
such as the Berezinskii-Kosterlitz-Thouless transition as well as quantum transitions involving a change in electronic band topology.
It is expected that gapped topological states of matter are generically robust 
to weak many-body interactions 
that do not close the single particle gap \cite{ schnyder_classification_2008, chiu_classification_2016,qi_topological_2011}. However, 
stronger interactions might lead to closing of the gap or to
collective modes induced by interactions, such as excitons, which can become unstable and drive symmetry breaking.
Landau's concept of a Fermi liquid (FL) with sharp electron-like quasiparticles 
is also often inadequate to describe strongly interacting fluids which may not exhibit 
long-lived electronic quasiparticles. This can happen for systems near quantum critical points or in systems which have both disorder and
strong correlations. Partial progress in understanding such non-Fermi liquids (nFLs) has come from constructing exactly solvable models like 
the Sachdev-Ye-Kitaev (SYK) model in the large-$N$ limit \cite{Song2017, Haldar2018, Chowdhury2018,Chowdhury2021}. 

In light of the above efforts, it is worthwhile to study the interplay between topological 
transitions and strong correlations.
A partial exploration of this idea was discussed in \cite{zhang_topological_2018}, where the SYK interaction was shown to 
renormalize the location of topological phase transitions between Chern insulator and trivial insulator phases
and to render the topological gap less stable to nonzero temperature.
Here, we explore interacting variants of two-orbital models constructed by Venderbos, Hu, and Kane \cite{venderbos_higher_2018} (abbreviated below to  `VHK') to describe Chern insulators
with different angular momentum for orbitals which leads to bands with higher Chern numbers. 
The trivial and higher Chern insulators in these models are separated by a quadratic band touching critical point which is protected
by crystalline $C_n$ rotational symmetry. 
A repulsive interorbital Hubbard interaction is marginally relevant at this band touching point in two dimensions, and drives the formation of nematic order which breaks the $C_n$ rotational symmetry \cite{sun_topological_2009}. 
In fact, this physics of topological transition with quadratic band touching 
and emergent nematicity was first studied for $C\!=\!2$ Chern insulators with $C_6$ symmetry 
by Cook et. al. \cite{cook_emergent_2014}. The broader interplay of interactions and band touching points
in driving diverse symmetry broken phases has received considerable attention in previous work
\cite{vafekInteractingFermionsHoneycomb2010, maciejkoFieldTheoryQuantum2013, Janssen2017,Setty2022, Setty2023}.

In our work, we consider a variant of these models with 
random multi-orbital SYK interactions in the large-$N$ limit
and study anomalous Hall transport across the phase diagram. 
Since the SYK interactions are inherently random, the symmetries of the full Hamiltonian, discussed below in more detail, are `average symmetries' which are present upon disorder averaging.  
While this is obvious for lattice symmetries like translations, we show that it can also be true for on-site symmetries like charge conjugation.
We note that the role of average symmetries, both unitary and anti-unitary, on non-interacting as well as interacting symmetry-protected
topological (SPT) phases continues to be of interest
\cite{Disorder_Fu_PRL2012,Disorder_Mong_PRL2012,Disorder_Akhmerov_PRB2014,Disorder_Beenakker_PRB2015,Disorder_Ma2023}. 
Here, we discuss examples where the microscopic SYK model may not possess such symmetries, but they appear in the disorder averaged action.
We show that the phase diagram of the VHK model with random interactions includes, in addition to the correlated trivial and Chern insulator phases, a nematic phase driven by `exciton condensation'.
The main difference with respect to the Hubbard model is that the disorder averaged SYK interaction appears
to be irrelevant at the band touching point, and one needs to exceed a critical SYK coupling to drive nematic order.
If we start from the interacting single-site limit, our model corresponds to a topological lattice generalization of a coupled two-dot SYK model
 \cite{maldacena_eternal_2018, klebanov_spontaneous_2020, sahoo_traversable_2020}, which hosts a `wormhole' corresponding to spontaneous breaking
of an axial $U(1)$ symmetry. On the lattice, this
`wormhole' exciton condensate transmutes into a nematic state which breaks discrete
lattice rotational symmetry. The lattice model does not retain this $U(1)$ symmetry, although it does develop a nonzero exciton expectation value, so we retain this language.
\par
The VHK type models break time-reversal ${\cal T}$, and do not need charge conjugation symmetry ${\cal C}$ 
or chiral symmetry ${\cal S}$, and could thus be viewed as an enrichment of class $\mathbf{A}$ of the topological periodic 
table upon including crystalline $C_n$ rotational symmetry.
For simplicity, however, we endow it in our work with an additional charge conjugation ${\cal C}$ symmetry with ${\cal C}^2=+1$, as done in the instances of these tight-binding models studied by VHK. This ensures that the model has bulk insulating
phases except at critical points corresponding to topological phase transitions. The
Hamiltonians we study are enrichments of class $\mathbf{D}$, with a particle-hole symmetric spectrum and an integer Chern invariant. \par
A  distinct viewpoint on this class $\mathbf{D}$ VHK model is obtained by making a particle-hole transformation on one orbital. 
The resulting non-interacting phase diagram then describes, at the level of Bogoliubov deGennes (BdG) mean field theory, 
a gapped trivial superconductor and a gapped topological superconductor separated by a quadratic band touching critical point. 
In this language, our work effectively explores the impact of strong SYK interactions on this BdG Hamiltonian, 
and one of the phases we uncover corresponds to a strongly interacting nematic superconductor. \par
Recent experiments on hexagonal Bernal bilayer graphene as well as magic angle twisted bilayer graphene that have shown the wealth of
phases arising in Dirac and quadratic band touching systems
\cite{Yan2010, Kang2018, Hejazi2019,Kwan2020,cao2021, delabarrera2022,zhou2022,zhangEnhancedSuperconductivitySpin2023, holleis2023}. 
These materials show evidence for nematic order in the normal and superconducting states,
exhibit anomalous Hall effects, and display non-Fermi liquid behavior at
high temperature via $T$-linear resistivity \cite{polshyn2019}. 
These experiments serve as inspiration for our exploration of the interplay between topology and interactions.

\section{Hamiltonian and symmetries} \label{model}

\subsection{Noninteracting Model}
We consider the two-dimensional (2D) VHK model of a Chern insulator
\begin{equation}
H_{0} =\sum_{\bk} \varphi_{\bk}^{\dg} h_{\bk}^\pdg \varphi_{\bk}^\pdg, \quad \varphi_{\bk}=\left(\begin{array}{l}
c_{\bk,1} \\
c_{\bk,2}
\end{array}\right)
\end{equation}
Here $c_{\bk,\s}$ are annihilation operators for electrons in 
orbitals $\sigma=1,2$ with momentum $\bk$.
For obtaining topological bands with Chern number $C = \ell$, the orbitals should have at least relative angular momentum $\ell$. 
For $C = \pm 3$, this corresponds to atomic $(s, f)$ orbitals 
with $\varphi^T_{\bk} = ({s}_{\bk}, {f}_{\bk})$; we will focus on this case in the
body of the paper.
The case with $s$-$p$ or $s$-$d$ orbital pairs for $\ell=1,2$ respectively
are discussed in Appendix E.

We study the triangular lattice
with $D_{6h}$ symmetry, choosing $s$ to transform as $A_{1g}$ and $f$ to transform
as $B_{1u}$. We consider the Hamiltonian
matrix
\begin{equation}
h^\pdg_{\bk}=\varepsilon^\pdg_{\bk} \t^\pdg_z+\Delta^\pdg_1 
\lambda_{\bk}^{(1)} \t^\pdg_y+\Delta^\pdg_2 \lambda_{\bk}^{(2)} \t^\pdg_x
\end{equation}
where $\lambda_{\bk}^{(1)}$ and 
$\lambda_{\bk}^{(2)}$ are basis functions for $B_{1u}$ and $B_{2u}$ representations,
while $\varepsilon^\pdg_{\bk}$ is the $A_{1g}$ basis function.

For our model, ${\cal T}^2 = +1$, and time-reversal corresponds simply to complex conjugation ${\cal K}$.
Let us define mirror operations ${\cal M}_y: (k_x,k_y) \to (k_x,-k_y)$ and ${\cal M}_x: (k_x,k_y) \to (-k_x,k_y)$.
We can then examine all the symmetries of this model.

(1) $C_6$: $\varphi_\bk \to \tau_z \varphi_{{\cal R}\bk}$
where rotation ${\cal R} \equiv {\cal R}_{2\pi/6}$.
Thus $C_6$ symmetry implies $\tau_z h_\bk \tau_z = h_{{\cal R}\bk}$, which is
satisfied since $\varepsilon_\bk = \varepsilon_{{\cal R}\bk}$, while $\lambda^{(i)}_{\bk} 
= -\lambda^{(i)}_{{\cal R}\bk}$. 
We later deduce nematic order in the interacting model by examining whether $\Delta_{\bk} = i \Delta^\pdg_1 \lambda_{\bk}^{(1)} + \Delta^\pdg_2 
\lambda_{\bk}^{(2)}$ develops additional terms that are not odd under $C_6$ rotation. For the specific case
of $\ell=3$, the $C_6 \to C_3$ nematic symmetry breaking we observe later can also be thought of as inversion
breaking; for $\ell=2$, nematic order does not imply inversion breaking.

(2) ${\cal M}_y {\cal T}$: $\varphi_\bk \to {\cal K} \varphi_{\bk'}$, where $\bk' = {\cal M}_y \cdot (-\bk)$.
We then expect $h^\pdg_{\bk} \to h^*_{\bk'}$. This is satisfied since $\varepsilon_{\bk'} = \varepsilon_\bk$,
$\lambda_{\bk'}^{(1)} = -\lambda_{\bk}^{(1)}$, $\lambda_{\bk'}^{(2)} = \lambda_{\bk}^{(2)}$, and $\tau_y^* = -\tau_y$.

(3) ${\cal M}_x {\cal T}$: $\varphi_\bk \to {\cal K} \tau_z \varphi_{\bk'}$, where $\bk' = {\cal M}_x \cdot (-\bk)$.
We then expect $h^\pdg_{\bk} \to \tau_z h^*_{\bk'} \tau_z$. This is  satisfied since $\varepsilon_{\bk'} = \varepsilon_\bk$,
$\lambda_{\bk'}^{(1)} = \lambda_{\bk}^{(1)}$, $\lambda_{\bk'}^{(2)} = -\lambda_{\bk}^{(2)}$, and $\tau_y^* = -\tau_y$.

(4) Charge conjugation ${\cal C}$: $\varphi^\dg_\bk \to \varphi^T_{-\bk} \tau_x$.
This leads to $h^\pdg_{\bk} \to - \tau_x h^T_{-\bk} \tau_x$. 
This symmetry is obeyed by our Hamiltonian.

Finally, we note that the combination of charge conjugation and $C_6$ symmetry does not permit a term in $h_\bk$
which is proportional to the identity matrix.

Working with nearest and next-nearest neighbor hoppings, we choose basis
functions,
\begin{eqnarray}
\varepsilon_\bk &=& \frac{t}{2} (z- 2 \sum^3_{i=1} \cos k_i) - \delta \\
\lambda_{\bk}^{(1)}&=&\sum_{i=1}^3 \sin k_i \\ 
\lambda_{\bk}^{(2)}&=&\frac{-1}{3 \sqrt{3}} \sum_{i=1}^3 \sin \left(k_i-k_{i+1}\right),
\end{eqnarray}
where the coordination number $z=6$, 
and $\delta$ tunes the band topology.
We have defined $k_i = \bk.\hat{a}_i$ where $\hat{a}_i = \cos\theta_i \hat{x} + \sin\theta_i \hat{y}$
with $\theta_i = (i-1) 2 \pi/3$.

\begin{figure}
\centering
\includegraphics[width=0.48\textwidth]{ 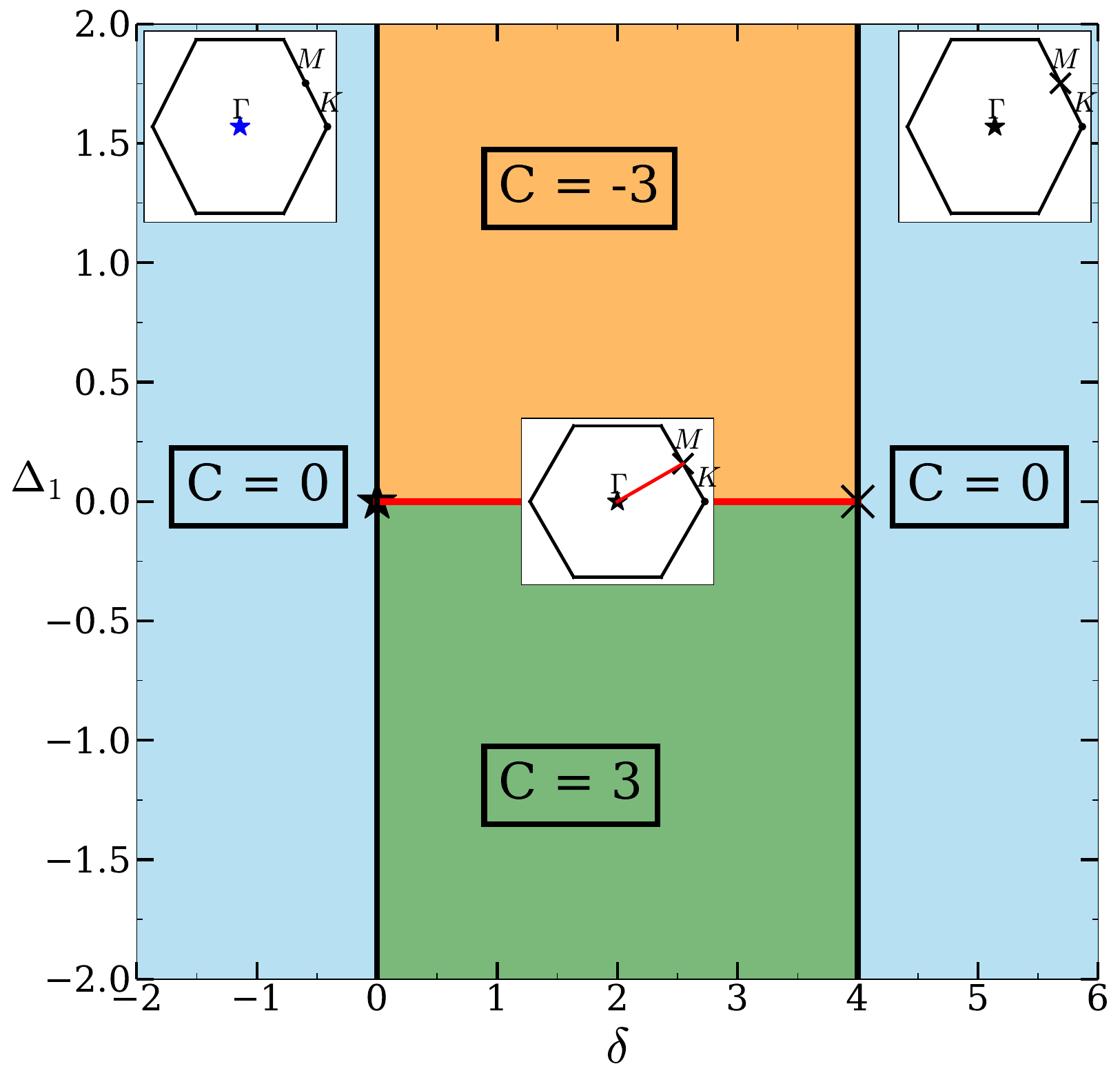}
\caption{Non-interacting phase diagram of the $\ell\!=\!3$ VHK model upon tuning $\delta$ and $\Delta_1$ for
fixed $t\!=\!\Delta_2\!=\!1$.  
The $C = 0 \mapsto \pm 3 $ transition at $\delta = 0$ occurs when the gap closes at the $\mathbf{\Gamma}$ point.
The $C =\pm 3 \mapsto 0 $ transition at $\delta = 4$ occurs when the gap closes at $\mathbf{M}$ points.
The $C = 3 \mapsto -3 $ transition occurs via band touchings at generically incommensurate momenta along $\mathbf{\Gamma}-\mathbf{K} $ high-symmetry line. }
\label{fig:phasediag_vhk}
\end{figure}

\begin{figure*}
\centering
\includegraphics[width=1.0\textwidth]{ 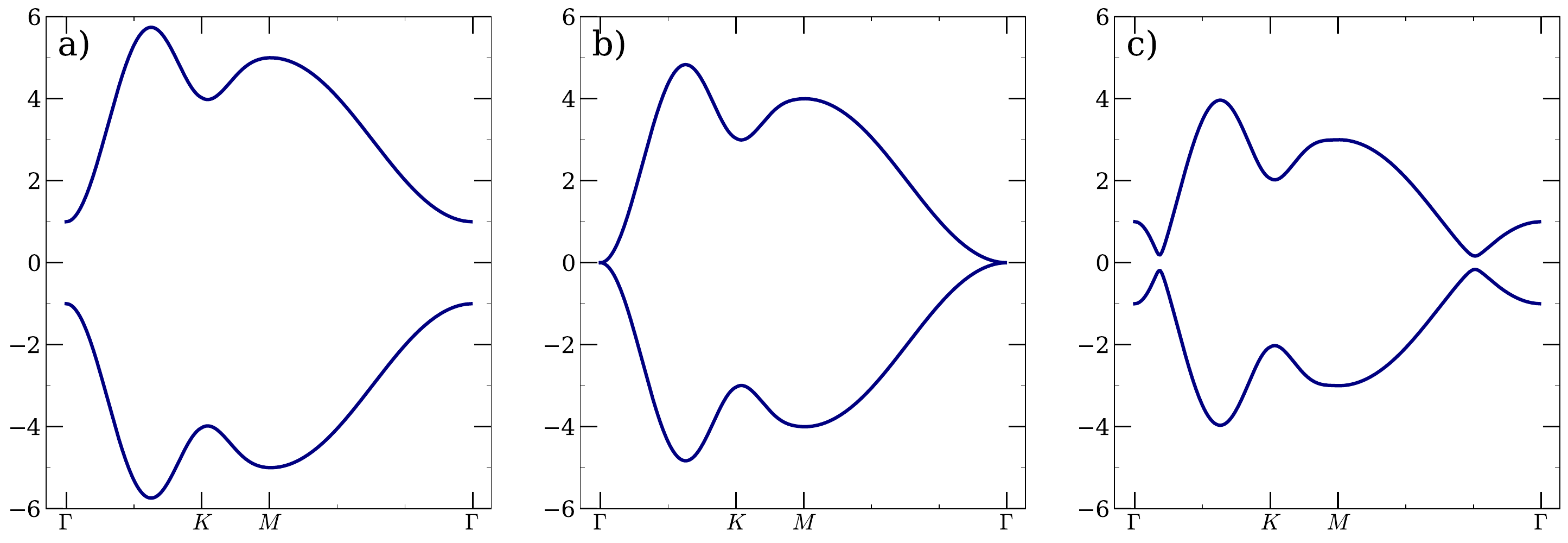}
\caption{The non-interacting band-structure of the $\ell=3$ VHK lattice model. Here,  energies are in units of $t$, and we have set $\Delta_1 = \Delta_2=t$.
The bands are shown for (a) the trivial regime with $\delta = -1$, (b) the quadratic band-touching point for $\delta = 0$, and (c) the topological regime for $\delta = 1$. }
\label{fig:bandstructure}
\end{figure*}
Fig.~\ref{fig:phasediag_vhk} shows the phase diagram of this non-interacting VHK model when we fix $t=\Delta_2=1$ and tune $\delta$ and $\Delta_1$. We
find trivial band insulators with $C=0$ as well as quantum anomalous Hall insulators with $C=\pm 3$.
The band touching points at the topological phase transitions are also indicated.
The $C=0 \to \pm 3$ transitions are driven by a quadratic band touching point with winding number $\pm 3$ at the $\mathbf{\Gamma}$ point or by three winding number $\pm 1$ band touchings at the $M$ points.
The $C=3 \leftrightarrow -3$ transitions proceed via six Dirac band touchings located at generic incommensurate momenta located
along the $\mathbf{\Gamma}-\mathbf{K}$ lines. 
Below we discuss in detail the impact of interactions on the $C= 0 \to 3$ transition at $\Delta_1=1$ as we tune $\delta$. The band structure as we go across this transition is shown in Fig.\ref{fig:bandstructure}. 
Band touchings at the $\mathbf{M}$ points or incommensurate momenta might nucleate more complex broken symmetries associated with spatially modulated orders; given the
numerical complexity of this problem, we defer this exploration to future work.
However, a discussion of models with $\ell=1,2$ with $\mathbf{\Gamma}$ point band touchings is given in Appendix E, mainly to show that the phenomenology in those cases is similar to the $\ell=3$ model explored here.

\subsection{SYK interactions }

We next switch on SYK interactions on each site by generalizing each orbital $\s$ to have additional `flavor' indices, and consider 
the Hamiltonian
\begin{equation}
\begin{gathered}
H_{\rm SYK} =  \sum_{\br}\sum_{ijkl}\sum_{{\s_i}}J_{ijkl}^{\s_1\s_2\s_3\s_4}(\br)
  O^{\s_1\s_2\s_3\s_4}_{ijkl}(\br) \\
O^{\s_1\s_2\s_3\s_4}_{ijkl}(\br) = c_{\br,\s_1,i}^{\dagger} c_{\br,\s_2,j}^{\dagger} c^\pdg_{\br,\s_3,k} 
  c^\pdg_{\br,\s_4,l}.
\end{gathered}
\end{equation}
Here $\s_{1-4}$ take on values $1,2$.
The flavor indices $i,j,k,l$ take on values $1\ldots N$, and we will be interested in the limit $N \to \infty$.
We denote $\bar{\s} = \neg \s$.
The random couplings $J_{ijkl}$ are (anti)symmetrized to obey $J^{\s_1\s_2\s_3\s_4}_{ijkl}=-J^{\s_2\s_1\s_3\s_4}_{jikl}=-J^{\s_1\s_2\s_4\s_3}_{ijlk}$.
Hermiticity
fixes $J^{\s_1\s_2\s_3\s_4}_{ijkl} = (J^{\s_4\s_3\s_2\s_1}_{lkji})^*$. 
The spatial symmetries $C_6$, ${\cal M}_x {\cal T}$, and ${\cal M}_y {\cal T}$ do not act on the flavor indices.
Furthermore, for the random SYK couplings, these symmetries (as well as translations) will be assumed to be present 
only on average, so they will not relate couplings on different sites taken into each other under the symmetry operation.

We consider the following interaction Hamiltonian for the main body of the text
\begin{equation}
H_{\rm SYK} = \sum_{\br, ijkl,\s} \left[J_{ijkl} (\br)
  \left( O^{\s\s\s\s}_{ijkl}(\br) + \a O^{\s\bar{\s}\s\bar{\s}}_{ijkl}(\br) \right)\right]
  \label{interaction}
\end{equation}
where $J_{ijkl}$ is a complex random variable, independent of $\s$, which is sampled from a Gaussian 
distribution with $\langle J_{ijkl}(\br)\rangle=0$, and
$\langle J^*_{ijkl}(\br) J^\pdg_{ijkl}(\br')\rangle =\delta_{\br \br^\prime} J^2/(2N)^{3}$. 
We impose that $J^{1111}_{ijkl} =  \a J_{ijkl}^{1212} $ with a real proportionality constant $\alpha$, motivated by potential physical realizations of this model \cite{sahoo_traversable_2020, lantagne-hurtubise_superconducting_2021}. 
At the single site level, $H_{\rm SYK}$ preserves a local $U(1) \otimes U(1)$ symmetry that spontaneously breaks into a single $U(1)$ symmetry \cite{klebanov_spontaneous_2020, sahoo_traversable_2020}.
Related models have received considerable attention \cite{zhouTunnelingEternalTraversable2020,  plugge_revival_2020, antoniniHolographicBoundaryStates2021, lantagne-hurtubise_diagnosing_2020, garcia-garciaPhaseDiagramTwosite2021, caiNonHermitianQuantumSystem2022} for their insight into the relation between quantum information and spacetime geometries. 
In the present formulation, this exciton condensation transition corresponds to 
an axial U(1) symmetry breaking mechanism. In an alternative particle-hole transformed formulation,
this is a global U(1) symmetry breaking which leads to superconducting condensate \cite{lantagne-hurtubise_superconducting_2021}. The lattice extension of this is discussed in Appendix B, where again, although the $U(1)$ symmetry has already been broken by the lattice pairing terms, the interactions develop a novel nematic superconducting order parameter. \par
Here, we explore the phase diagram and transport 
for the model $H_{0}+H_{\rm SYK}$ as we tune band topology and interaction 
parameters $J,\alpha$. 
This microscopic interaction $H_{\rm SYK}$ has none of the point group symmetries or the charge conjugation symmetry
of the non-interacting model. However, disorder averaging via the replica trick,
and taking the large $N$ limit, we find that all symmetries, ${\cal C}$, $C_6$, ${\cal M}_x {\cal T}$, and
${\cal M}_y {\cal T}$ emerge as ``average symmetries'' of $H_{0}+H_{\rm SYK}$ (Appendix A). 
\par

\section{Large-$N$ solution}

\subsection{Dyson equation and self-energy}

Starting from the Hamiltonian $H_0+H_{\rm SYK}$, we derive the disorder averaged effective $G \Sigma$ action using the replica trick. 
The details of such a procedure have become common \cite{Chowdhury2021}, so we defer some pertinent details to Appendix A and proceed by writing out the Dyson equations for the local Green function and the self-energy. 
\begin{equation}
    G_{\s\s^\prime}(i \w)=\int \frac{\text{d} \bk }{(2 \pi)^2}\ (i \w - h(\bk) - \Sigma(i \w))^{-1}_{\s \s^\prime}
\end{equation}
\begin{equation}
\begin{gathered}
\Sigma_{\s \s^\prime}(\t)   = -J^2 \Bigg( G^2_{\s \s^\prime}(\t) G_{\s^\prime \s}(-\t) \\
+ \a G_{\s \s^\prime}(\t)\Big(G_{\s \bar{\s}^\prime}\left(-\t \right) G_{\bar{\s}\s}\left(\t \right)+G_{\bar{\s} \s^\prime}\left(-\t\right) G_{\s^\prime\bar{\s}}\left(\t \right) \Big) 
\\ + \frac{\alpha^2}{2} G_{\bar{\s}^\prime\bar{\s} }(-\t)\big(G_{\s \bar{\s}^\prime}(\t) G_{\bar{\s} \s^\prime}(\t) +  G_{\s \s^\prime}(\t)G_{\bar{\s}\bar{\s}^\prime}(\t) \big) \Bigg)
\end{gathered}
\end{equation}
Solving these equations self-consistently also gives us the free-energy presented in Appendix A, Eq.\ \ref{Free_Energy}, and we use it to compute order parameters and to study the phase transitions mentioned below. 
\subsection{Anomalous Hall Transport}
In order to determine the topological phases, we compute $\sigma_{xy}$.
 The anomalous Hall conductivity is derived from the zero-momentum paramagnetic current-current correlation function
 \begin{equation}
 \s^{\mu\nu}(\w)=- \lim_{\eta \rightarrow  0}\frac{1}{\w} K^{\mu\nu}(\mathbf{Q} = \mathbf{0},\w+i \eta).
\end{equation}
 where the correlation function for uniform frequency is given by \cite{acheche_orbital_2019}
 \begin{equation}
 \begin{gathered}
     K^{\m\n}( i \w_n) = e^2\sum_{i \nu_n} \sum_{\bk }
     \text{Tr}\left(\partial_{\m}h(\bk)\vdot G(\bk, i \nu_n)\vdot \right. \\
    \left. \partial_{\n}h(\bk)\vdot G(\bk, i \w_n + i \nu_n)\right) 
 \end{gathered}
 \end{equation}
 This quantity can be computed upon analytic continuation to real frequency. 
 Here, since we have a matrix expression for $G$ with general non-zero off-diagonal entries, the spectral function takes a more general form. 
\begin{equation}
     A(\bk, \w) = \frac{i}{2 \pi }\left(G^R(\bk,\w) - G^R(\bk,\w)^\dagger\right)
 \end{equation}
In order to perform numerical computations , we analytically continue the self-consistent equations  \ref{eq:selfcon2} themselves, from the Matsubara frequencies ($i\w_n$) to the real frequency line ($\w\in\mathbb{R}$) following the method presented in Ref.\ \cite{Maldacena2016, sahoo_traversable_2020} which is detailed in Appendix C. 
The ability to solve self-consistent equations directly on the real-time axis is a major motivator for the study of the SYK interactions.  
The real-time self-consistent equations provides the matrix retarded Green's functions $G^{R}(\w)$ and the matrix retarded self-energies $\Sigma^{R}(\w)$. Additional details regarding the calculation can be found in Appendix D.

\section{Phase diagram} 
\begin{figure}
\centering
\includegraphics[width=0.475\textwidth]{ 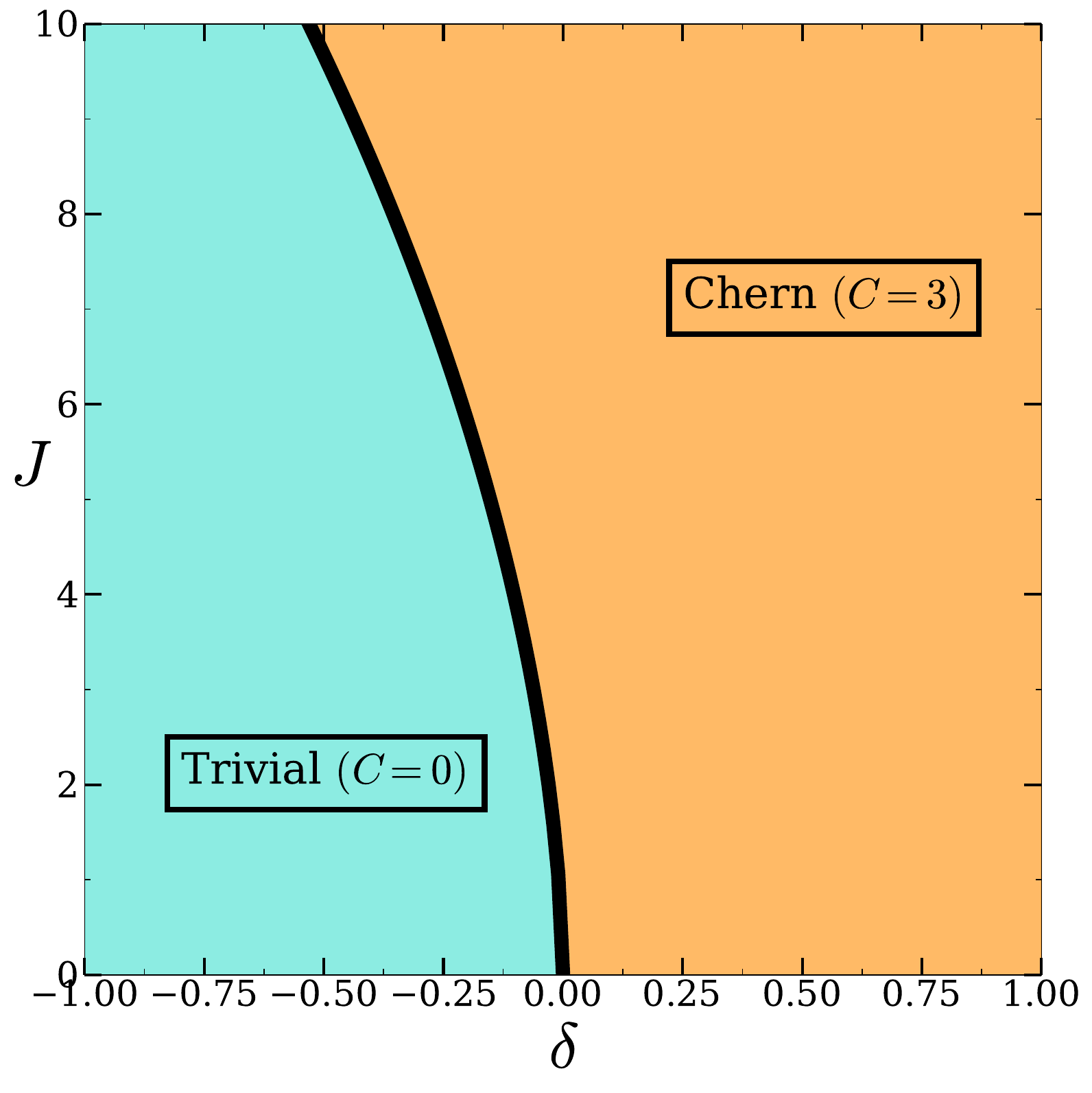}
\caption{Zero temperature phase diagram as a function of tuning mass parameter $\delta$ and the effective SYK interaction strength $J / t$ for $\a= 0$.
(The numerical solution is obtained at $T/t = 10^{-2}$.)
The topological transition is impacted by $J$ via a renormalization of $\delta$. Both phases, the trivial insulator
and Chern insulator, preserve crystalline $C_6$ rotational symmetry.} 
\label{fig:topphasediag}
\end{figure}
\label{phase transitions}
At low-temperatures, these SYK interactions construct a rich phase-diagram. 
We restrict our interest to low-temperatures for several reasons. 
The quantized Chern number and topological phase transitions are strictly defined only in the zero-temperature limit $T=0$ 
\cite{schnyder_classification_2008, chiu_classification_2016}. 
The higher-dimensional SYK model also demonstrates a FL-nFL crossover at $T \gtrsim t^2/J$, which can be demonstrated from a power-counting argument that the fermion-hopping term is a relevant perturbation to the four-fermion incoherent interaction \cite{Song2017}.
This means in the region where the topological aspects of this model are apparent, there remains a well-defined coherent quasiparticle picture. 
The hopping $t$ also serves to reduce the density of states which reduces the interaction-induced scattering rate. 
This crossover does not exclude the possibility that interactions at finite coupling strength can have dramatic effects on the low-temperature physics, as we demonstrate here. 
Finite temperature behaviour can be seen through the anomalous Hall response.  
We begin by considering the fate of the topological phase transition without the inter-orbital coupling, i.e with $\alpha=0$, 
and then ask about the phase diagram when $\alpha \neq 0$ is switched on.
\subsection{$ \a = 0$ Phase Diagram}
The non-interacting lattice model we study undergoes a topological phase transition from $C = 0$ to $C = 3$ 
at $\delta = 0$ and another which returns it to the trivial phase at $\delta = 4$.
This transition occurs due to band-inversions of the quadratic band-touching points (QBTP). 
This QBTP occurs at the $\mathbf{\Gamma}$ point at $\delta = 0$ in all of the constructed lattice models. The point is isotropic in $\bk$ space. 
We focus our investigation about this point because the QBTP nature of point has an enhanced density of states. 
The recombination band-touching point at $\delta = 4$ generically occurs at non-zero $\bk$, although the interaction phenomenology remains similar.  \par

In order to determine the effect of correlated interactions on this topological transition, we compute the anomalous Hall conductivity $\s_{xy}$ using the approach detailed in Section III B. 
This transition also holds for the $\mathcal{C}-$conjugated superconducting model.
We focus our attention on the Chern insulator because the Chern number can be extracted through $\s_{xy} =C/ 2 \pi $, which is no longer possible in a superconducting setting where charge is not conserved.  
$C$ refers to the homotopy invariant Chern number that distinguishes the system from a topologically trivial and non-trivial state.  \par  

The anomalous Hall conductivity also serves as a more robust measure of topological invariants than previously defined zero-temperature invariants that implicitly rely on single-particle wave-functions or propagators and are limited to zero temperature \cite{Wang2010, Gurarie2011, Zhao_2023, Setty2023}.  
Fig. \ref{fig:topphasediag} demonstrates the effect on this phase boundaries upon tuning $J$ for $\a= 0$. When $\a= 0$, the effect of the SYK interaction within each orbital serves to shift the transition boundary. 
This increases the topological non-trivial region. 
The shifted topological phase diagram can be explained through an effective Hamiltonian \cite{cook_emergent_2014,zhang_topological_2018}; 
\begin{equation}
h_{\mathrm{eff}}(\bk)=-G_R^{-1}(\w=0, \bk).
\end{equation}
The $\mathcal{C}$ symmetry indicates that for a diagonal $G(\w = 0, \bk)$, the matrix structure of $\Sigma(0) \propto \Sigma_{11}(0) \t_z$ is such that $\Sigma_{11}(0) = - \Sigma_{22}(0) $. This implies that there is an effective renormalization of $\delta$ in this picture, given by 
\begin{equation}
    \delta \rightarrow \delta +  \frac{\Sigma_{11}(0)}{t} .
\end{equation}
This renormalization shifts the bare value of $\delta$ at which the band touching occurs and the resulting topological band inversion takes place
\cite{zhang_topological_2018}. \par
\UseRawInputEncoding
When $\a\neq 0$, there is not a simple matrix form of $\Sigma(0)$, although for $\Sigma_{11,22}$, $J$ is effectively renormalized by $J_{\mathrm{eff}} = J(1+ \alpha^2/2)$, which in turn contributes to the renormalization of $\delta$. 

\subsection{$\a \neq 0$ Phase Diagram }
\label{phase_diagram}
\begin{figure}
\centering
\includegraphics[width=0.5\textwidth]{ 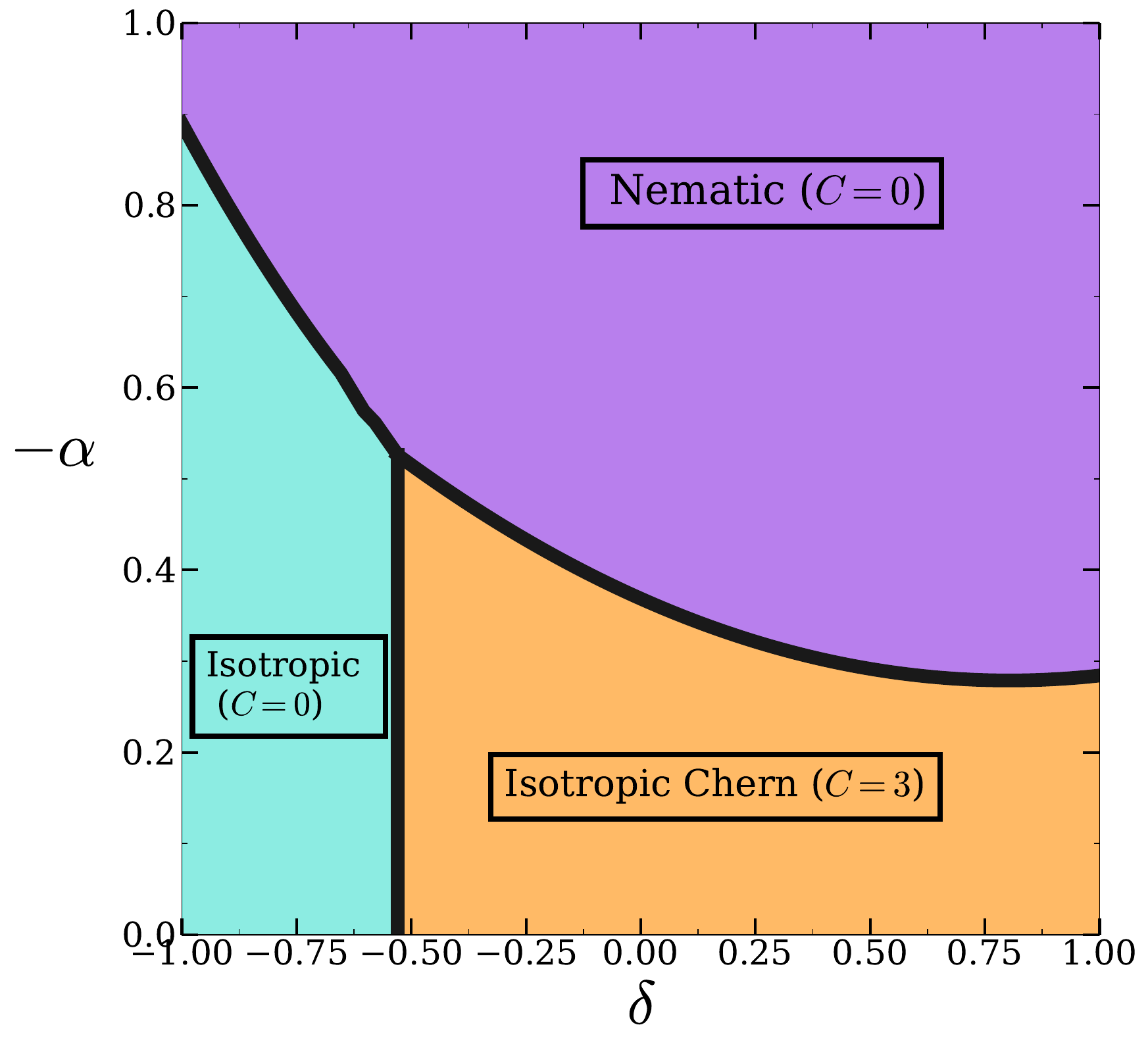}
\caption{The zero temperature phase diagram at strong coupling $J/t = 10$ (our numerical solutions are for $T/t = 10^{-2}$). 
The critical coupling constant $-\alpha_c$ for inducing nematic order decreases as the band gap is reduced, but reaches its minimum finite value 
deeper in the topological phase when the orbital populations are most mixed so that the inter-orbital interaction can play an important role.} 
\label{fig:nematicphase}
\end{figure}
As described in Section \ref{model}, the non-interacting Hamiltonian is invariant under lattice point group symmetries, such as rotation. 
The introduction of local interaction terms given by Eq.\ \ref{interaction} respect this symmetry in the Hamiltonian on average. 
Previous mean-field studies of a marginal Hubbard interaction \cite{cook_emergent_2014, venderbos_higher_2018}, revealed exciton ordering due to the increased density of states at the quadratic-band touching points. 
This transition is due to an induced exciton self-energy $\left(\Sigma_{\s \bar{\s}}(i \w)\right)$.
The formation of such particle-hole pairs in the $\ell= 0$ channel and their `condensation' shifts the 
coupling of the orbitals as $\Delta_{\bk} \rightarrow \Delta_{\bk} + \Delta_0$,
which breaks the rotational symmetry due to the spontaneously generated momentum independent contribution. 
This is distinct from nematic transitions where anisotropic orbitals (with preferred directions) develop spontaneous population imbalances \cite{hardy_nematic_2023}. \par
The order parameter of the transition in our model can be extracted as 
\begin{equation}
    \! \Delta_0 \!=\! \frac{1}{2N} \int  \frac{\mathrm{d}\bk}{(2\pi)^2} \ \langle \varphi^\dagger_{\bk} \t_x \varphi^\pdg_{\bk} \rangle \!=\! \int \frac{\mathrm{d}\bk}{(2\pi)^2} \  G_{12}(\bk, \t = 0) 
\end{equation}

An intra-orbital SYK interaction does not lead to such spontaneous symmetry breaking. 
Hence, we proceed by asking whether or not the inter-orbital $\a$ term induces such exciton formation. 
The 0+1D model spontaneously develops a nonzero exciton order parameter for $\a< 0$ and for $\a> 4$ \cite{sahoo_traversable_2020}. 
One can show that there exists a duality between $\a = 0, 4$ equivalent only to renormalizing $J_{\rm eff}$ \cite{sahoo_traversable_2020}. 
For this reason, we explore the stability of this $\a = 0$ transition in the higher-dimensional $2+1$ D model.
In this $2+1$ D model, the thermal nematic transition is typically continuous, 
although the low-temperature phase transition upon tuning $\alpha$ is first order, consistent with the $0+1$D model. \par

\par

Fig. \ref{fig:nematicphase} demonstrates the phase diagram of tuning $\delta$ and $\alpha$ at a large value of $J = 10$. We focus on the regime of the phase diagram for $\alpha < 0$ since the 2+1D theory displays no spontaneous symmetry breaking for $\alpha > 0$, at least
for $0 < \a <4$, as in the 0+1D theory. 
We have not explored larger values $\alpha > 4$ in this work since our primary interest was to
study the physics of the quadratic band touching point. A similar phase diagram exists for fixed $\a< 0 $ and tuning $J$. We
discuss this phase diagram in Appendix.\ \ref{fig:tuningJ} and Fig.\ \ref{fig:Jphase}.
 The starting point of our analysis is a strongly interacting theory, dominated by $J$.
 $\a $ induces an instability from this strongly-interacting theory. 
This is in sharp contrast to mean-field Hubbard results, where $U$ is a marginal interaction that the non-interacting Fermi liquid is unstable to. 

 \par

The transition requires nonzero occupation in both orbitals in order to spontaneously break rotational symmetry. 
This means that the topological phase, with large band-inversion, is more susceptible to the transition since the bands are naturally inverted after a Chern transition, leading to occupation in both the bands. 
The required orbital mixing decreases the critical $\alpha$ as a function of $\delta$. \par
\begin{figure}
\centering
\includegraphics[width=0.45\textwidth]{ 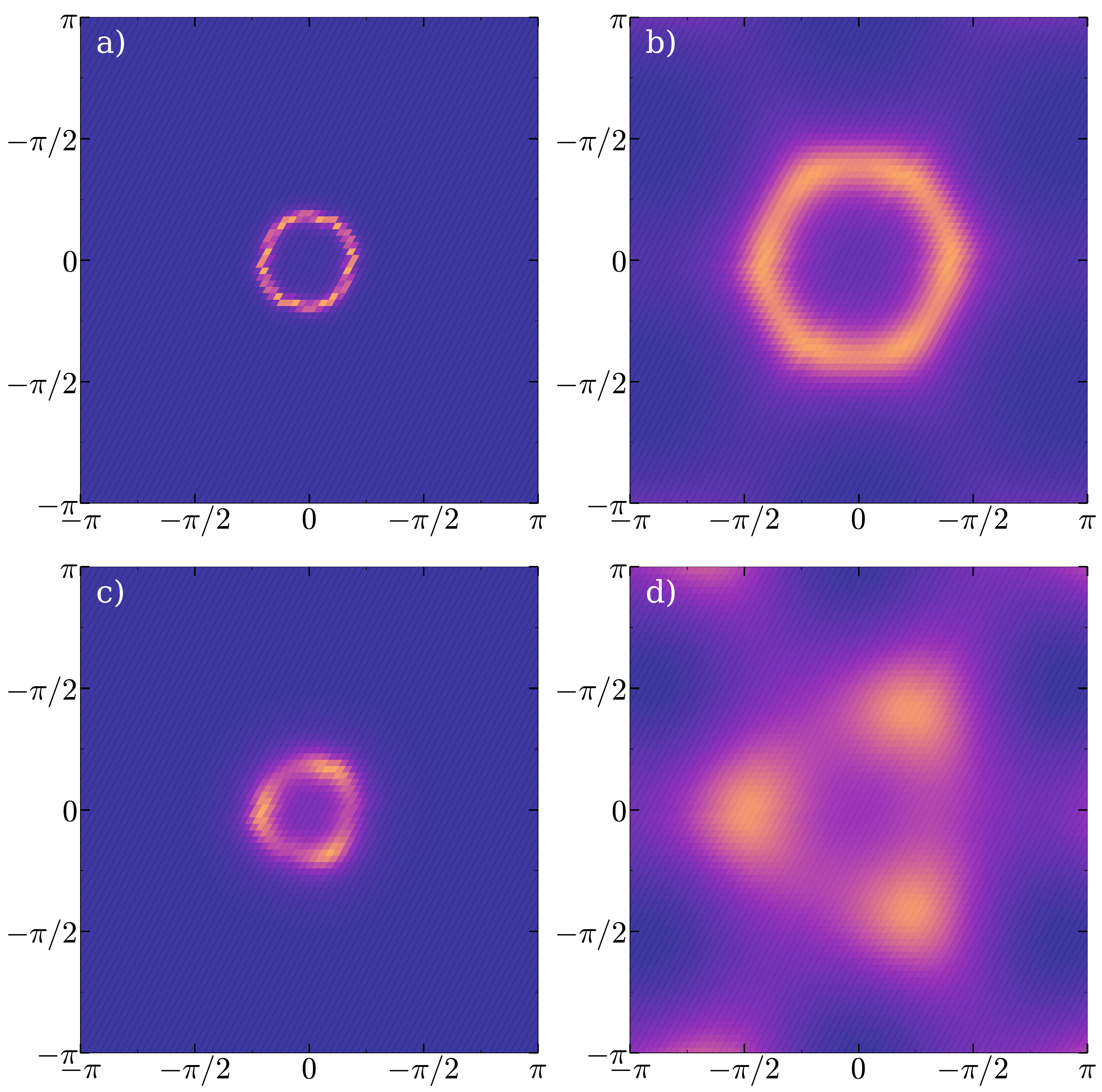}
\caption{Low energy frequency integrated spectral weight $\int_0^{\Omega_c} \operatorname{Tr}A(\bk, \w)$, with $\Omega_c = 0.05$ centered at the $\Gamma$ point. We consider the
non-interacting theory a) without and c) with a mean-field nematic order parameter $\Delta_0$. We show the  interacting theory with 
b) $\alpha = 0$ without and d) $\alpha = -0.5$ with spontaneous symmetry breaking. All figures are shown at $\delta = 0.35$. The 
interacting spectral functions demonstrate a  clear nematic symmetry breaking down to $C_3$.}
\label{fig:spectralfunctions}
\end{figure}

We demonstrate nematic order through the low energy frequency integrated spectral weight $\int_0^{\Omega_c} \operatorname{Tr}A(\bk, \w)$ in Fig. \ref{fig:spectralfunctions}. with $\Omega_c = 0.05$. Here we exhibit the effects of intra-orbital SYK interactions on the spectral functions without symmetry breaking, a mean-field example of the nematic order parameter, and the fully interacting theory with inter-orbital interactions. The intra-orbital interactions broaden the spectral functions, although they retain the symmetry. The inter-orbital interactions induce spontaneous symmetry breaking that reduces the symmetry to $C_3$. 
 \par
\begin{figure*}[h]
\centering
\includegraphics[width=1.0\textwidth]{ 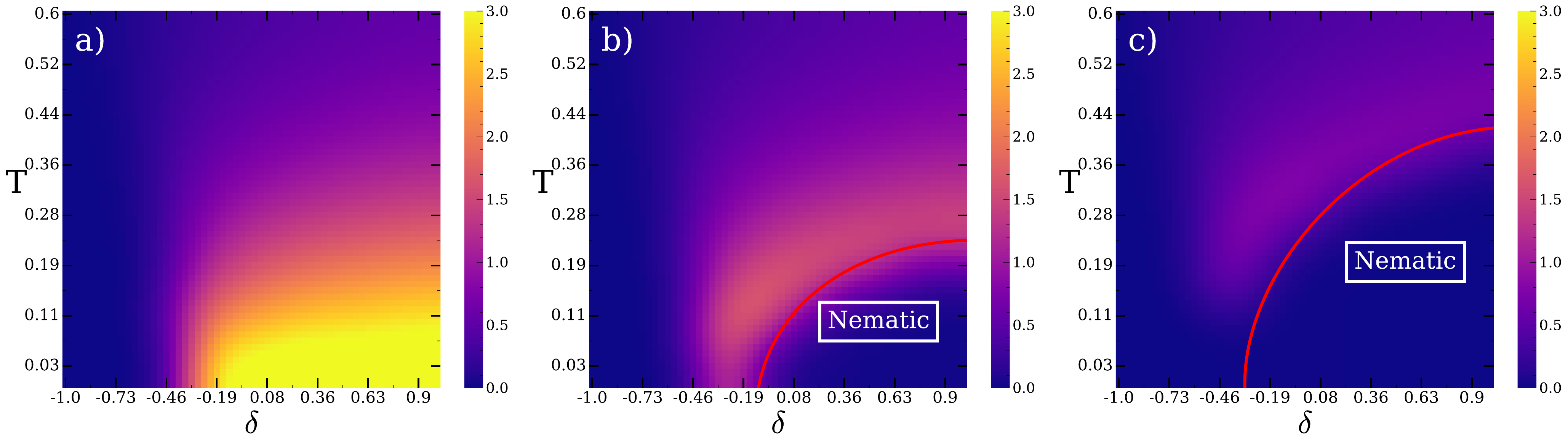}
\caption{ A plot of the anomalous Hall conductivity $\s_{xy}$ as a function of the mass $\delta$ and temperature $T$ for the inter-orbital correlation a) $ \a = -0.1$ , b) $ \a = -0.4$, and c) $\a= -0.5$. 
The Chern transition has been shifted away from $\delta = 0$ by intra-orbital $J/t = 10$ with $\Delta =1$.
Figure 5.b) demonstrates that a clear nematic dome suppresses the onset of quantized anomalous Hall conductivity given by a Chern number. 
The reduction of the anomalous Hall conductivity for the larger $\a$ occurs at a much higher temperature.  
The red lines indicate nematic order from free-energy calculations. At sufficiently low $T$, the critical $\delta$ becomes $T$ independent and ends in a first-order transition point.  }
\label{fig:hall}
\end{figure*}

The effect of these interactions on the anomalous Hall conductivity, $\s_{xy}$, is explored in Fig.\ \ref{fig:hall} for finite temperatures. 
In the topological phase, $\s_{xy}$ is unquantized at high temperatures but approaches a quantized value ($C = 3$) as we lower temperature, typical of a topological phase. 
A nonzero $\sigma_{xy}$ persists at high-temperature, even in the regime with incoherent quasiparticles, similar to observations in previous work \cite{zhang_topological_2018}. 
We attribute this to local contributions from fermions which hop around elementary triangular plaquettes of the lattice.
This can contribute to the anomalous Hall response even in the absence of a Fermi surface or well-developed topological bands as has been shown, for instance, in strongly disordered ferromagnetic semiconductors \cite{Burkov_DisorderedAHE2003}.
We note that the impact of the SYK interaction $J$ has shifted the trivial-topological phase boundary away from $\delta = 0$; this is also visible at finite temperatures.
Increasing $J$ however serves to broaden the spectral functions which reduces the Hall gap. The interactions therefore destabilize the quantized Hall conductivity to temperature fluctuations. This provides an explanation for the observed behaviour in \cite{zhang_topological_2018}.
 At larger $\d$, as we go deeper into the topological phase, a nonzero $\s_{xy}$
persists to even higher temperature. 
This behaviour contrasts with the fact that, for sufficiently large $\a$ and low temperature, inter-orbital interactions suppresses the anomalous Hall conductivity, and result in a trivial $C = 0$ state. Upon increasing $\a$, the anomalous Hall conductivity therefore decreases much more rapidly upon heating.

\section{Outlook} \label{outlook}

In this work, we have considered a strongly correlated model where topological phase transitions and Chern numbers can be computed without reference to band-structure and wave-functions or much of the framework for non-interacting or weakly-interacting topological insulators. 
This work extends the notions of average symmetries that have been considered in relation to topological phases. 
We have demonstrated the role that strong interactions have on the excitations of a general class of these insulators and how they shape both symmetry-broken and topological phase transitions. 
The exciton pairing interaction at $\a\neq 0$ induces a spontaneous rotational broken symmetry from $C_6 \mapsto C_3$. 
This induced nematic order shifts the phase-boundaries of the topological phase transition and at sufficiently large interactions, preempts the topological phase transition. 
This model can also serve to describe nematic superconducting order parameters in topological superconductors. 
As compared to marginal mean-field results with Hubbard-like interactions, the density population of both orbitals play a dramatic role in the transition.

Going beyond the regime of a clean hopping model with disordered interactions, if we include weak randomness in the hopping Hamiltonian or go beyond large-$N$ for the interactions, based on an Imry-Ma argument \cite{ImryMa_PRL1975}, 
we expect the nematic phase to be strictly destroyed since there will be local random fields which break $C_6$ symmetry.
However, large nematic domains might nevertheless persist at weak disorder, and the topological phases and anomalous Hall transport are expected to survive.

An interesting future avenue is motivated by previous work \cite{Haldar2018}. In $d = 2$, power-counting arguments demonstrate that $\a$ interactions would only be relevant for quite flat bands with dispersion $\varepsilon(\bk) \propto |\bk|^p$ $p > 6$.
However, for $p \geq 3$ a distinct non-Fermi liquid state persists to $T = 0$, different than the SYK solution, despite the interactions being irrelevant. 
The nature of this intermediary regime deserves further attention. \par
One can also ask a related question based on work on nematic order within 3D quadratic band-touching systems \cite{Moon2013,Janssen2017, Fei2018}. Short-range interactions in that case are RG irrelevant, although long-range interactions are strongly relevant and lead to non-Fermi liquid behaviour.
The Yukawa-SYK model in $2$D also has a critical bosonic propagator which also leads to non-Fermi liquid behaviour. \cite{esterlisCooperPairingIncoherent2019,wangSolvableStrongCouplingQuantumDot2020, esterlisLargeTheoryCritical2021}. 
The role of such critical behaviour in the current model is worthy of exploration. There are a few interesting variants of the model that could also be explored.
If the random interaction respected translational symmetry, the disorder average would induce a momentum dependence in $\Sigma(\bk, \w)$ \cite{Chowdhury2018}. 
One could also study doped versions of the model away from half-filling to examine the effects of Berry curvature and interactions in a partially filled metallic Chern band.  

Future directions of this work can also turn towards the aspects of the coupled SYK model that remain unexplored in higher dimensional problems. 
Considerable recent interest in this class of coupled SYK orbitals demonstrates a wide variety of exotic phenomena, such as revival dynamics, wormholes, and the teleportation of quantum information \cite{maldacena_eternal_2018, lantagne-hurtubise_diagnosing_2020, sahoo_traversable_2020}. 
These aspects of the model have even been probed through quantum computing simulations, although the physics observed in this simplified model remain unclear \cite{jafferis2022, kobrin2023}.
The role that non-trivial band-structure and topology has on such phenomena remains undetermined.  

\begin{acknowledgments}
We thank Arijit Haldar, Andre-Marie Tremblay, Stephan Plugge, and Etienne Lantagne-Hurtubise for useful discussions. 
We acknowledge support from the Natural Sciences and Engineering Research Council (NSERC) of Canada. A. Hardy acknowledges support from a NSERC Graduate Fellowship (PGS-D). 
Numerical computations were performed on the Niagara supercomputer at the SciNet HPC Consortium and the Digital Research Alliance of Canada.

\end{acknowledgments}
\bibliography{apssamp}
\appendix
\onecolumngrid
\clearpage

\section{Self Consistent Solutions for a Coupled Correlated SYK model}\label{Appendix:SelfCons}
We begin with the Hamiltonian 
\begin{equation}
H=\sum_{\bk} \varphi_{\bk}^{\dagger} h_{\bk} \varphi_{\bk}, \quad \varphi_{\bk}=\left(\begin{array}{l}
c_{\bk, 1} \\
c_{\bk, 2}
\end{array}\right),
\end{equation}
with the interaction terms 
\begin{equation}
    H^{\rm intra}_{J}= \sum_{\br}\sum_{ijkl}\sum_{\s} J_{ijkl}(\br)
  c_{\br,\s,i}^{\dagger} c_{\br,\s,j}^{\dagger} c^\pdg_{\br,\s,k} 
  c^\pdg_{\br,\s,l} 
\end{equation}
\begin{equation}
H^{\rm inter}_{\alpha} =\a\sum_{\br}\sum_{ijkl}\sum_{\s} J_{i j  k l}(\br)c_{\br, \s, i}^{\dagger} c_{\br,\bar{\s}, j}^{\dagger} c^\pdg_{\br,\s, k} c^\pdg_{\br,\bar{\s}, l}.
\end{equation}
The interaction terms can be rewritten using fermion quadlinears as 
\begin{equation}
\begin{gathered}
    H_{\textbf{SYK}} = 
    H_J + H_\a \equiv \sum_{\br}\sum_{ijkl} J_{ijkl}(\br) \mathcal{O}_{ijkl}(\br, \t) = \sum_{\br}\sum_{ijkl}\sum_{\s} J_{ijkl}(\br) (O^{\s \s \s \s }_{ijkl}(\br, \t)+ \a O^{\s \bar{\s}\s\bar{\s}  }_{ijkl}(\br, \t)) \\
    = \sum_{\br}\sum_{i<j, k<l}\sum_{\s} J_{ijkl}(\br) \left(4 O^{\s \s \s \s }_{ijkl}(\br, \t)+ 2\a\left(  O^{\s \bar{\s}\s\bar{\s}  }_{ijkl}(\br, \t) + O^{\s \bar{\s}\bar{\s}\s}_{ijkl}(\br, \t) \right) \right) 
    \end{gathered}
\end{equation}
where 
\begin{equation}
   O^{\s_1 \s_2 \s_3 \s_4}_{ijkl}(\br, \t) = c_{\br, \s_1, i}^\dagger(\t) c_{\br,  \s_2, j}^\dagger(\t) c_{\br, \s_3, k}(\t) c_{\br,\s_4, l}(\t) ,
\end{equation}
and the second equality in (A4) is for the restricted sum $i < j, k < l$. This means that the disorder averaging is over independent indices. 
Here $\bar{\s}$ refers to the opposite orbital and $\s^\prime$ refers to either the same or opposite orbital. 
After following the replica trick \cite{Chowdhury2021}, we have the disorder-averaged partition function as 
\begin{equation}
\begin{gathered}
\overline{Z} = \int  \prod_s \mathcal{D}[\bar{c}, c]
\exp \left[ -  \int \text{d} \t_1  \sum_{\br, \br^\prime}\sum_{\s, \s'}\sum_{i} \bar{c}_{\br, \s,i} \left((\partial_{\t_1} - \mu) \delta_{\br \br'}\delta_{\s \s^\prime} + h_{\s \s^\prime} (\br - \br^\prime) \right) c_{\br',\s,i} \right. \\
 \left. - \frac{J^{2}}{16 N^3} \int \text{d} \t_1 \text{d}\t_2  \left(\sum_{\br}\sum_{i<j,k<l} \mathcal{O}_{ijkl}( \br, \t_1)\mathcal{O}_{lkji}(\br, \t_2)  \right)\right].
 \label{action}
\end{gathered}
\end{equation}
This partition function can be rewritten in terms of fermion bilinears $G$ and $\Sigma$ by applying a Lagrange multiplier constraint which is detailed in \cite{Chowdhury2021}.
The effective saddle-point action in the $N \rightarrow \infty$ limit is given as 
\begin{equation}
\begin{gathered}
-\frac{1}{N} S[G, \Sigma]= \int \frac{\text{d} \bk }{(2 \pi)^2}\ \text{d} \t \ln \operatorname{det}\left(\left(\partial_\t-\mu\right) \delta_{\s \s^\prime }-h_{\s \s^\prime }(\bk)-\Sigma_{\s \s^\prime}(\tau) \right) \\ 
+\int \text{d} \t_1 \text{d} \t_2\left\{\sum_{\s \s^\prime} \Sigma_{\s \s^\prime}\left(\t_1, \t_2\right) G_{ \s^\prime \s}\left(\t_2, \t_1\right)+\frac{J^2}{4}\left(\sum_{\s \s^\prime}  G_{\s \s^\prime}\left(\t_1, \t_2\right)^2 G_{ \s^\prime \s}\left(\t_2, \t_1\right)^2\right.\right.\\
+2 \alpha\left[G_{11}\left(\t_1, \t_2\right) G_{11}\left(\t_2, \t_1\right)+G_{22}\left(\t_1, \t_2\right) G_{22}\left(\t_2, \t_1\right)\right]\left[G_{12}\left(\t_2, \t_1\right) G_{21}\left(\t_1, \t_2\right)+G_{21}\left(\t_2, \t_1\right) G_{12}\left(\t_1, \t_2\right)\right] \\
+\alpha^2\left[G_{11}\left(\t_2, \t_1\right) G_{22}\left(\t_2, \t_1\right) G_{11}\left(\t_1, \t_2\right) G_{22}\left(\t_1, \t_2\right)+G_{12}\left(\t_2, \t_1\right) G_{21}\left(\t_2, \t_1\right) G_{11}\left(\t_1, \t_2\right) G_{22}\left(\t_1, \t_2\right)\right.\\
\left.+G_{11}\left(\t_2, \t_1\right) G_{22}\left(\t_2, \t_1\right) G_{21}\left(\t_1, \t_2\right) G_{12}\left(\t_1, \t_2\right)+G_{12}\left(\t_2, \t_1\right) G_{21}\left(\t_2, \t_1\right) G_{21}\left(\t_1, \t_2\right) G_{12}\left(\t_1, \t_2\right)\right]\Bigg) \Bigg\}.
 \label{action_2}
\end{gathered}
\end{equation}
The corresponding Dyson equations for $\Sigma_{\s\s}$ or $\Sigma_{\s\bar{\s}} $ is given as (with $\t$-invariance)

\begin{equation}
    G_{\s\s^\prime}(i \w)=\int \frac{\text{d} \bk }{(2 \pi)^2}\ (i \w - h(\bk) - \Sigma(i \w))^{-1}_{\s \s^\prime}
\end{equation}
where 
\begin{equation}
\begin{gathered}
\Sigma_{\s\s}(\t)  = -J^2 \Bigg( G^2_{\s\s}(\t)G_{\s\s}(-\t)+ \a G_{\s\s}(\t)\Big(G_{\s\bar{\s}}\left(-\t \right) G_{\bar{\s} \s}\left(\t \right)+G_{\bar{\s} \s}\left(-\t\right) G_{\s\bar{\s}}\left(\t \right) \Big)
\\ + \frac{\alpha^2}{2} G_{\bar{\s}\bar{\s}}(-\t)\left( G_{\s\s}(\t)G_{\bar{\s}\bar{\s}}(\t)+  G_{\s\bar{\s}}(\t)G_{\bar{\s} \s}(\t)\right) \Bigg)
\end{gathered}
\end{equation}
\begin{equation}
\begin{gathered}
\Sigma_{\s\bar{\s}}(\t)  = -J^2 \Bigg( G^2_{\s\bar{\s}}(\t)G_{\bar{\s} \s}(-\t)+ \a G_{\s\bar{\s}}(\t)\Big(G_{\s\s}(-\t) G_{\s\s}(\t )+G_{\bar{\s}\bar{\s}}(-\t) G_{\bar{\s}\bar{\s}}\left(\t \right) \Bigg) 
\\ + \frac{\alpha^2}{2} G_{\s\bar{\s}}(-\t)\left(G_{\s\s}(\t)G_{\bar{\s}\bar{\s}}(\t)+  G_{\s\bar{\s}}(\t)G_{\bar{\s} \s}(\t)\right) \Bigg).
\end{gathered}
\end{equation}

The condensed general expression is given as 

\begin{equation}
\begin{gathered}
\label{eq:selfcon2}
\Sigma_{\s \s^\prime}(\t)   = -J^2 \Bigg( G^2_{\s \s^\prime}(\t) G_{\s^\prime\s}(-\t) 
+ \a G_{\s \s^\prime}(\t)\Big(G_{\s \bar{\s}^\prime}\left(-\t \right) G_{\bar{\s}^\prime \s}\left(\t \right)+G_{\bar{\s} \s^\prime}\left(-\t\right) G_{\s^\prime\bar{\s}}\left(\t \right) \Big) 
\\ + \frac{\alpha^2}{2} G_{\bar{\s}^\prime\bar{\s} }(-\t)\left(G_{\s \bar{\s}^\prime}(\t) G_{\bar{\s} \s^\prime}(\t) +  G_{\s \s^\prime}(\t)G_{\bar{\s},\bar{\s}^\prime}(\t) \right) \Bigg) ,
\end{gathered}
\end{equation}
The corresponding free-energy density is given by 
\begin{equation}
\begin{gathered}
    f = \sum_n \ln{\det(G^{-1})} + \sum_{\s \s^\prime} \int_0^\beta \text{d} \t  \Bigg\{ \Sigma_{\s\s^\prime} ( \tau)  G_{\s\s^\prime} (\beta -\tau) -   \frac{J^2}{4} \Bigg( G^2_{\s\s^\prime} ( \tau)  G^2_{\s^\prime \s} (\beta -\tau) \\
    +  \a\Big(G_{\s\bar{\s}} ( \tau) G_{\bar{\s}^\prime \s} (\tau) G_{\s\s^\prime}(\beta - \tau) G_{\s \bar{\s}^\prime} (\beta -\tau)  + G_{\bar{\s} \s^\prime} ( \tau) G_{\sigma^\prime \bar{\s}}( \tau) G_{\bar{\s} \s^\prime} (\beta -\tau)G_{\s \s^\prime} (\beta -\tau)\Big) \\
     \frac{\alpha^2}{2} \Big( G_{\s\bar{\s}}(\beta - \tau) G_{\bar{\s}^\prime\bar{\s} }(\beta - \tau) G_{\s \bar{\s}^\prime} (\tau) G_{\bar{\s} \s^\prime}(\tau) + G_{\s\bar{\s}}(\beta - \tau) G_{\bar{\s}^\prime\bar{\s} }(\beta - \tau) G_{\s \s^\prime} (\tau) G_{\bar{\s}\bar{\s}^\prime}(\tau) \Big) \Bigg) \Bigg\}
     \label{Free_Energy}
\end{gathered}
\end{equation}
\section{Superconducting model}
We can recast the above model in BdG version by making a particle hole transformation on one orbital, $c_{i 2}^{\dagger} \leftrightarrow c_{i 2}$. This 
leads to
\begin{equation}
H_{\rm BdG} =\sum_{\bk} \varphi_{\bk}^{\dagger} h^\pdg_{\bk} \varphi^\pdg_{\bk}, \quad \varphi_{\bk}=\left(\begin{array}{l}
c_{\bk, 1} \\
c^\dagger_{-\bk, 2}
\end{array}\right)
\end{equation}
Such a model describes a class ${\mathbf D}$ mean field superconductor with inter-orbital pairing and broken time-reversal
symmetry.
This Hamiltonian can describe a transition between a trivial superconductor and a topological superconductor,
which is driven by tuning the chemical potential. \par
The derivation for the interaction nematic superconducting model follows Appendix B upon applying this particle hole transformation and sending $\a \mapsto - \a$. The equivalent symmetry is now $J^{\s\s\s\s}_{ijkl} = J^{\bar{\s}\bar{\s}\bar{\s}\bar{\s}}_{klij} = (J^{\bar{\s}\bar{\s}\bar{\s}\bar{\s}}_{ijkl})^*$

Our disorder averaged partition function is still given by Eq.\ \ref{action}. The quadlinears over the restricted flavour indices are now given by
\begin{equation}
\mathcal{O}_{ijkl}(\br, \t) = 4 O^{1111}_{ijkl}(\br, \t)+4 O^{2222}_{klij}(\br, \t)+ 2\alpha(  O^{1212}_{ilkj}(\br, \t) + O^{2121}_{kjil}(\br, \t) + O^{2112}_{ljki}(\br, \t) + O^{1221}_{ikjl}(\br, \t))  
\end{equation}

An equivalent procedure to above gives \cite{lantagne-hurtubise_superconducting_2021}
\begin{equation}
\begin{gathered}
-\frac{1}{N} S[G, \Sigma]= \int \frac{\text{d} \bk }{(2 \pi)^2}\ \text{d}\t \ln \operatorname{det}\left(\left(\partial_\t-\mu\right) \delta_{\s \s^\prime }-h_{\s \s^\prime }(\bk)-\Sigma_{\s \s^\prime }(\tau)\right) \\ 
+\int d \t_1 d \t_2\left\{\sum_{\s} \Sigma_{\s \s}\left(\t_1, \t_2\right) G_{\s \s}\left(\t_2, \t_1\right)+ \sum_{\s} \Pi_{\s \bar{\s}} F_{\bar{\s}\s} + \right. \\
\frac{J^2}{4}\left(\sum_{\s} G_{ \s \s}\left(\t_1, \t_2\right)^2 G_{ \s \s}\left(\t_2, \t_1\right)^2 +F_{ \s \s}\left(\t_1, \t_2\right)^2  F_{ \s \s}\left(\t_2, \t_1\right)^2 \right.\\
-2 \alpha\left[G_{11}\left(\t_1, \t_2\right) G_{11}\left(\t_2, \t_1\right)+G_{22}\left(\t_1, \t_2\right) G_{22}\left(\t_2, \t_1\right)\right]\left[F_{12}\left(\t_2, \t_1\right) F_{21}\left(\t_1, \t_2\right)+F_{21}\left(\t_2, \t_1\right) F_{12}\left(\t_1, \t_2\right)\right] \\
+\alpha^2\left[G_{11}\left(\t_2, \t_1\right) G_{22}\left(\t_2, \t_1\right) G_{11}\left(\t_1, \t_2\right) G_{22}\left(\t_1, \t_2\right)+F_{12}\left(\t_2, \t_1\right) F_{21}\left(\t_2, \t_1\right) G_{11}\left(\t_1, \t_2\right) G_{22}\left(\t_1, \t_2\right)\right.\\
\left.+G_{11}\left(\t_2, \t_1\right) G_{22}\left(\t_2, \t_1\right) F_{21}\left(\t_1, \t_2\right) F_{12}\left(\t_1, \t_2\right)+F_{12}\left(\t_2, \t_1\right) F_{21}\left(\t_2, \t_1\right) F_{21}\left(\t_1, \t_2\right) F_{12}\left(\t_1, \t_2\right)\right]\Bigg)\Bigg\}
 \label{action_3}
\end{gathered}
\end{equation}
This gives equivalent Dyson equations to \ref{eq:selfcon2} upon the substitutions $\a\mapsto - \a$, $G_{\s, \bar{\s}} \mapsto F_{\s, \bar{\s}}$, and $\Sigma_{\s, \bar{\s}}  \mapsto \Pi_{\s, \bar{\s}}$
\section{Analytically Continued Dyson Equations}
The self-consistent Dyson equations can be analytically continued in order to be solved in real time explicitly \cite{Maldacena2016, Banerjee2017, sahoo_traversable_2020, hardy_nematic_2023}. Following that procedure provides the self-consistent real-frequency $\Sigma(\w)$.
\begin{equation}
\begin{gathered}
\Sigma_{\s\s}(\w)   = -i J^2 \int d \w e^{i \w t} \Bigg( (n_{\s\s}^{++})^2 n_{\s\s}^{--}  + (n_{\s\s}^{-+})^2 n_{\s\s}^{+-}  \\
+\a\Big(n_{\s\s}^{++} n_{\bar{\s} \s}^{++} n_{\s\bar{\s}}^{--}+n_{\s\s}^{-+} n_{\bar{\s} \s}^{-+} n_{\s\bar{\s}}^{+-} + n_{\s\s}^{++} n_{\s\bar{\s}}^{++} n_{\bar{\s} \s}^{--}+n_{\s\s}^{-+} n_{\s\bar{\s}}^{-+} n_{\bar{\s} \s}^{+-}) \Big)
\\  
+\frac{\alpha^2}{2} \Big(n_{\s\s}^{++}n_{\bar{\s}\bar{\s}}^{++} n_{\bar{\s}\bar{\s}}^{--}  + n_{\s\s}^{-+} n_{\bar{\s}\bar{\s}}^{-+}n_{\bar{\s}\bar{\s}}^{+-} +  n_{\s\bar{\s}}^{++}n_{\bar{\s} \s}^{++} n_{\bar{\s}\bar{\s}}^{--}+ n_{\s\bar{\s}}^{-+}n_{\bar{\s} \s}^{-+} n_{\bar{\s}\bar{\s}}^{+-}  \Big)\Bigg)
\end{gathered}
\end{equation}
and 
\begin{equation}
\begin{gathered}
\Sigma_{\s\bar{\s}}(\w)   = -i J^2 \int d \w e^{i \w t} \Bigg( (n_{\s\bar{\s}}^{++})^2 n_{\bar{\s} \s}^{--}  + (n_{\s\bar{\s}}^{-+})^2 n_{\bar{\s} \s}^{+-}  \\
\a\Big(n_{\s\bar{\s}}^{++} n_{\s\s}^{++} n_{\s\s}^{--}+n_{\s\bar{\s}}^{-+} n_{\s\s}^{-+} n_{\s\s}^{+-} + n_{\s\bar{\s}}^{++} n_{\bar{\s}\bar{\s}}^{++} n_{\bar{\s}\bar{\s}}^{--}+n_{\s\bar{\s}}^{-+} n_{\bar{\s}\bar{\s}}^{-+} n_{\bar{\s}\bar{\s}}^{+-}) \Big)
\\  
+\frac{\alpha^2}{2} \Big(n_{\s\s}^{++}n_{\bar{\s}\bar{\s}}^{++} n_{\s\bar{\s}}^{--}  + n_{\s\s}^{-+} n_{\bar{\s}\bar{\s}}^{-+}n_{\s\bar{\s}}^{+-} +  n_{\s\bar{\s}}^{++}n_{\bar{\s} \s}^{++} n_{\s\bar{\s}}^{--}+ n_{\s\bar{\s}}^{-+}n_{\bar{\s} \s}^{-+} n_{\s\bar{\s}}^{+-}  \Big)\Bigg)
\end{gathered}
\end{equation}

where we have defined the time-dependent occupation as 
\begin{equation}
    n^{s s^\prime}(t)=\int d \w A_{\s \s^\prime} \left(\w\right) n_{\rm F}\left(s\w\right) e^{i s^\prime \w t},
\end{equation}
and the spectral function $ A_{\s \s^\prime} \left(\w\right) $ is given from the generalized matrix expression 
\begin{equation}
     A( \w) = \frac{i}{2 \pi }\left(G^R(\w) - G^R(\w)^\dagger\right)
 \end{equation}
The equivalent simplification can be made as 
\begin{equation}
\begin{gathered}
\Sigma_{\s \s^\prime}(\t)  = -i J^2 \Bigg( \int d \w e^{i \w t} \Bigg( (n_{\s \s^\prime}^{++})^2 n_{\s^\prime \s}^{--}  + (n_{\s \s^\prime}^{-+})^2 n_{\s^\prime \s}^{+-} \\ +\a \Big(n_{\s \s^\prime}^{++} n_{\bar{\s}^\prime\s}^{++} n_{\s \bar{\s}^\prime}^{--}+n_{\s \s^\prime}^{-+} n_{\bar{\s}^\prime\s}^{-+} n_{\s \bar{\s}^\prime}^{+-} + n_{\s \s^\prime}^{++} n_{b,-a}^{++} n_{\bar{\s} \s^\prime}^{--}+n_{\s \s^\prime}^{-+} n_{\bar{s}^\prime,\bar{s}}^{-+} n_{\bar{\s} \s^\prime}^{+-}) \Big)
\\  + \frac{\alpha^2}{2} \Big(n_{\s \bar{\s}^\prime}^{++}n_{\bar{\s} \s^\prime}^{++} n_{\bar{\s}^\prime\bar{\s} }^{--}  + n_{\s \bar{\s}^\prime}^{-+} n_{\bar{\s} \s^\prime}^{-+}n_{\bar{\s}^\prime\bar{\s} }^{+-} +  n_{\s \s^\prime}^{++}n_{\bar{\s}\bar{\s}^\prime}^{++} n_{\bar{\s}^\prime\bar{\s} }^{--}+ n_{\s \s^\prime}^{-+}n_{\bar{\s}\bar{\s}^\prime}^{-+} n_{\bar{\s}^\prime\bar{\s} }^{+-}  \Big)\Bigg)
\end{gathered}
\end{equation}

\section{Anomalous Hall Conductivity Derivation }

While the calculation of the anomalous Hall conductivity in terms of spectral functions is in principle routine, we provide the detailed analysis as we were unable to find a suitable reference in the literature to our knowledge. 
In order to construct the current operator, we start with a general hopping Hamiltonian, \(h(\bk)\) and consider a Pierls substitution$ k_\mu \rightarrow k_\mu - eA_\mu$. We are interested in the uniform DC conductivity, so we restrict to the uniform, zero momentum field.
This gives the paramagnetic current operator  as \cite{Nourafkan2018}
\begin{equation}
    j_\mu(\mathbf{Q} = 0, i\w) =\frac{\delta S[A]}{\delta A_\mu} =  - e \sum_{\bk}\sum_{i\nu} \varphi_{\bk, i\nu}^\dag \left( \sum_\mu \pdv{h(\bk)  }{\bk_\mu}   \right) \varphi_{\bk, i\w+i\nu}\,.
\end{equation}
We compute the paramagnetic current-current correlation function \cite{Altland_Simons_2010}
\begin{equation}
    K^{\mu\nu}(\mathbf{0}, i\w) = \expval{j_{\mu}(\mathbf{0}, i\w) j_{\nu}(\mathbf{0}, -i\w)}
\end{equation}
such that the conductivity can be extracted from the correlator as
 \begin{equation}
 \label{cond defn}
 \s^{\mu\nu}(\w)=- \lim_{\eta \rightarrow  0}\frac{1}{\w} \operatorname{Im} (K^{\mu\nu}(\mathbf{0},\w+i \eta)).
\end{equation}
The  correlator in Matsubara frequency is given as \cite{acheche_orbital_2019,Acheche2020}
\begin{equation}
    K^{\m\n}( i \w) = e^2\sum_{i \nu_n} \sum_{\bk }
    \text{Tr}\left(\partial_{\m}h(\bk)\vdot G(\bk, i \nu)\vdot \partial_{\n}h(\bk)\vdot G(\bk, i \w + i \nu)\right)\,, 
    \label{correlation}
\end{equation}
where \(G\) is the Green's function. For an interacting case this expression does not admit a direct analytic continuation. Instead, we utilize the spectral representation of the Greens functions, perform the Matsubara sum, and then analytically continue to arrive at 
\begin{equation}
\begin{gathered}
K^{\m \n}(\w) = - e^2 \sum_{\bk} \operatorname{Tr}\left( \int d \varepsilon_1 d\varepsilon_2(\partial_{\m} h(\bk)\vdot A\left(\bk, \varepsilon_1\right) \vdot \partial_{\n} h(\bk)\vdot A\left(\bk, \varepsilon_2\right) )\right)  \frac{\left(n_F\left(\varepsilon_1\right)-n_F\left(\varepsilon_2\right)\right)}{\w -( \varepsilon_2 - \varepsilon_1)+i\eta}
\end{gathered}
\end{equation}
The trace argument is written as 
\begin{equation}
\begin{gathered}
T^{\m \n}(\bk;\varepsilon_1, \varepsilon_2) = \left(\partial_{\m} h(\bk)\vdot A\left(\bk, \varepsilon_1\right) \vdot \partial_{\n} h(\bk)\vdot A\left(\bk, \varepsilon_2\right) \right).
\end{gathered}
\end{equation}
This gives two terms for the imaginary part of the response, the real $T^{\m\n}$  contribution is 
\begin{equation}
\label{Real contri}
\begin{gathered}
\operatorname{Im}(K^{\m \n}(\w))_1 = - e^2 \pi\sum_{\bk} \operatorname{Tr}\left( \int d \varepsilon \operatorname{Re}(T^{\m \n}(\bk;\varepsilon, \varepsilon+\w) )\right)  \left(n_F\left(\varepsilon\right)-n_F\left(\varepsilon+\w\right)\right).
\end{gathered}
\end{equation}

The contribution from the imaginary part of $T^{\m\n}$ is 
\begin{equation}
\label{imag contri}
\begin{gathered}
\operatorname{Im}(K^{\m \n}(\w))_2 = - e^2 \sum_{\bk} \operatorname{Tr}\left(\int d \varepsilon_1 d\varepsilon_2 \operatorname{Im}(T^{\m \n}(\varepsilon_1, \varepsilon_2)) \right) \left(n_F\left(\varepsilon_1 \right)-n_F\left(\varepsilon_2\right) \right)  \mathcal{P}\left(\frac{1}{(\w - (\varepsilon_2 - \varepsilon_1)) }\right)\,,
\end{gathered}
\end{equation}
where \(\mathcal{P}\) refers to the principal value. The numerical implementation of Eq.\ \ref{imag contri} is
\begin{equation}
\begin{gathered}
\operatorname{Im}(K^{\m \n}(\w))_2 = - e^2 \sum_{\bk} \operatorname{Tr}\left(\int d \varepsilon_1 d\varepsilon_2 \operatorname{Im}(T^{\m \n}(\bk;\varepsilon_1, \varepsilon_2)) \right) \left(n_F\left(\varepsilon_1 \right)-n_F\left(\varepsilon_2\right) \right)  (\frac{(\w -( \varepsilon_2 - \varepsilon_1))}{(\w - (\varepsilon_2 - \varepsilon_1))^2 + \eta^2 })
\end{gathered}
\end{equation}
where $\eta$ is an infinitesimal value. This term is often neglected as most work deals with strictly real diagonal spectral functions and appears to be missing from the available literature.   \par
Using \eqref{cond defn}, we find that the first contribution to the DC limit of the conductivity through \eqref{Real contri} reduces to,
\begin{equation}
    \label{sigma contr 1}
    \s^{\m\n}(\w\rightarrow 0)_{1} =\pdv{}{\w} \operatorname{Im}(K^{\m \n}(\w))_1 \bigg|_{\w=0}  =  -e^2 \pi \sum_{\bk }\int_{-\infty}^{\infty} d\e_1\text{Re} \left( T^{\m\n}(\bk; \e_1, \e_1)\right)\cdot\frac{\partial n_F(\w)}{\partial\w}\bigg|_{\w=\e_1}.
\end{equation}
whereas \eqref{imag contri} contributes to the DC conductivity as
\begin{equation}
     \s^{\m\n}(\w\rightarrow 0)_{2} = \pdv{}{\w} \operatorname{Im}(K^{\m \n}(\w))_2 \bigg|_{\w=0} =  e^2  \int d \varepsilon_1 d\varepsilon_2 \sum_{\bk}\left(\operatorname{Im}T_{\mu \nu}(\bk;\varepsilon_1, \varepsilon_2)  \right) \left(n_F\left(\varepsilon_1 \right)-n_F\left(\varepsilon_2\right) \right) \frac{\eta^2 -\Delta \varepsilon_{2, 1}^2}{(\Delta \varepsilon^2_{2, 1} + \eta^2)^2}\,.
\end{equation}
We get the total DC conductivity as \(\s^{\m\n}(\w\rightarrow 0) = \s^{\m\n}(\w\rightarrow 0)_{1} + \s^{\m\n}(\w\rightarrow 0)_{2}\).
\section{Tuning $J$}
In Section \ref{phase_diagram}, we discuss the phase diagram for fixed $J$, while tuning $\alpha$. The physics are not fundamentally altered by tuning $J$ with fixed $\alpha$. For completeness however, we include such a phase diagram. The main difference is a consequence of the mass renormalization detailed in the Fig. \ref{fig:topphasediag}. For fixed $J = 10$, the intraorbital interaction renormalizes the mass, which shift the effective phase boundary to the left to more negative $\delta$ values. 
\label{fig:tuningJ}
\begin{figure*}
\includegraphics[width=1.0\textwidth]
{ 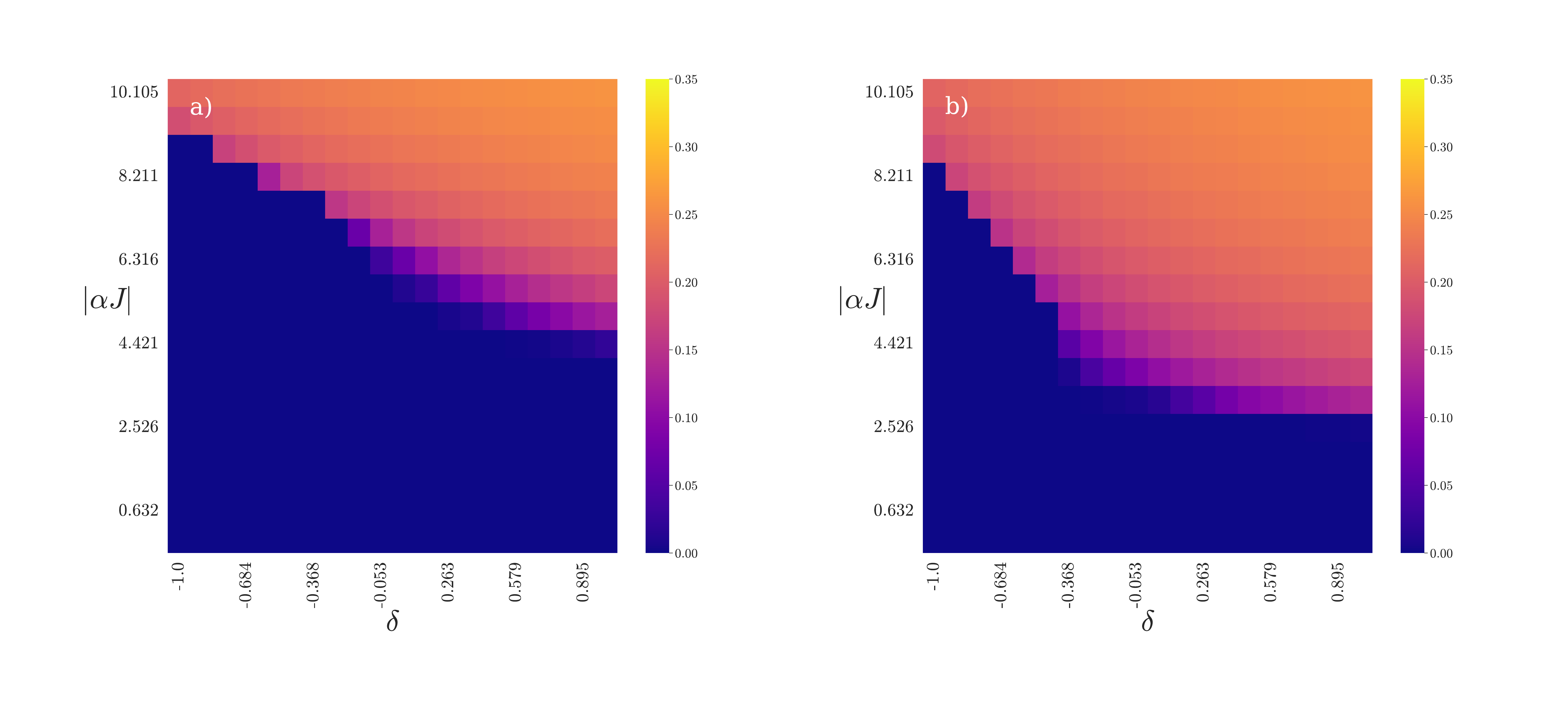}
\caption{The low-temperature phase diagram for $T/J = 1\times 10^{-3}$ and $\Delta = 1$ for fixed a) $\alpha = -1$ and b) $J = 10$. The color is the magnitude of our order parameter $\Delta_0$. The fixed $J$ plot demonstrates the effective shift of $\delta$ mentioned in the main text.  }
\label{fig:Jphase}
\end{figure*}

\section{Lower Angular Momentum Results}

In order to explore the relationship between the results for these phase transitions and the angular momentum $l$, we constructed lattice Hamiltonians with $l = 1$ and $l = 2$ corresponding to coupling with the $p$ and $d$ orbitals respectively \cite{venderbos_higher_2018} which will support Chern transitions of \(\pm 1\,,\pm 2\) respectively. There are also qualitative differences. All terms break $C_2$ symmetry for $C_6 \mapsto C_3$. the $l = 1$ and $l = 3$ nematic phases also break $\mathcal{I}$ as the orbital representation is $E_{1u}$ and $B_{1u}$, whereas the $l = 2$ case has an orbital representation of $E_{2g}$. This means that for $l =2$, $\mathcal{I}\Delta = \Delta$, whereas for $l = 1,3$    $\mathcal{I}\Delta = -\Delta$. We now consider the role that these symmetries have on our $l =1$ model. \par

(1) $C_6$: $\varphi_\bk \to \t_R\varphi_{{\cal R}\bk}$
where rotation ${\cal R} \equiv {\cal R}_{2\pi/6}$ and $\tau_R = \left((I + \t_z)/2 - \w (I - \t_z)/2 \right)$.
Thus $C_6$ symmetry implies $\tau^\dag_R h_\bk \tau_R = h_{{\cal R}\bk}$, which is
satisfied since $\varepsilon_\bk = \varepsilon_{{\cal R}\bk}$ and $\t^\dag_R \t_z \t_R = \t_z$, while $\t^\dag_R \t_+ \t_R = -\w\t_+$ , and  $\t^\dag_R \t_- \t_R = -\w^2\t_-$, while  $\lambda^{+}_{\bk} = -\w^2  \lambda^{+}_{{\cal R}\bk}
$ and  $\lambda^{-}_{\bk} = -\w  \lambda^{-}_{{\cal R}\bk}$.\par

(2) ${\cal M}_y {\cal T}$: $\varphi_\bk \to {\cal K} \varphi_{\bk'}$, where $\bk' = {\cal M}_y \cdot (-\bk)$.
We then expect $h^\pdg_{\bk} \to h^*_{\bk'}$. This is satisfied since $\varepsilon_{\bk'} = \varepsilon_\bk$, and 
$\lambda^{\pm} \rightarrow \lambda^{\pm}$. 

(3) ${\cal M}_x {\cal T}$: $\varphi_\bk \to {\cal K} \varphi_{\bk'}$, where $\bk' = {\cal M}_x \cdot (-\bk)$. This is identical to case (2)

(4) Charge conjugation ${\cal C}$: $\varphi^\dg_\bk \to \varphi^T_{-\bk} \tau_x$. This leads to
$h^\pdg_{\bk} \to - \tau_x h^T_{-\bk} \tau_x$. This symmetry is obeyed by our
Hamiltonian. \par

We now consider the role that these symmetries have on our $l =2$ model. \par

(1) $C_6$: $\varphi_\bk \to \t_R\varphi_{{\cal R}\bk}$
where rotation ${\cal R} \equiv {\cal R}_{2\pi/6}$ and $\tau_R = \left((I + \t_z)/2 + \w^2 (I - \t_z)/2 \right)$.
Thus $C_6$ symmetry implies $\tau^\dag_R h_\bk \tau_R = h_{{\cal R}\bk}$, which is
satisfied since $\varepsilon_\bk = \varepsilon_{{\cal R}\bk}$ and $\t^\dag_R \t_z \t_R = \t_z$, while $\t^\dag_R \t_+ \t_R = \w^2\t_+$ , and  $\t^\dag_R \t_- \t_R = \w\t_-$, while  $\lambda^{+}_{\bk} = \w  \lambda^{+}_{{\cal R}\bk}
$ and  $\lambda^{-}_{\bk} = \w^2  \lambda^{-}_{{\cal R}\bk}
$

(2) ${\cal M}_y {\cal T}$: $\varphi_\bk \to {\cal K} \varphi_{\bk'}$, where $\bk' = {\cal M}_y \cdot (-\bk)$.
We then expect $h^\pdg_{\bk} \to h^*_{\bk'}$. This is satisfied since $\varepsilon_{\bk'} = \varepsilon_\bk$, and 
$\lambda^{\pm} \rightarrow \lambda^{\pm}$. 

(3) ${\cal M}_x {\cal T}$: $\varphi_\bk \to {\cal K} \varphi_{\bk'}$, where $\bk' = {\cal M}_x \cdot (-\bk)$. This is identical to case (2)

(4) Charge conjugation ${\cal C}$: $\varphi^\dg_\bk \to \varphi^T_{-\bk} \tau_x$. This leads to
$h^\pdg_{\bk} \to - \tau_x h^T_{-\bk} \tau_x$. This symmetry is obeyed by our
Hamiltonian.

\par
\begin{equation}
h_{\bk}=\varepsilon_{\bk} \tau_{z}+\Delta \lambda_{\bk}^{l_{-}} \tau_{+}+\Delta^{*} \lambda_{\bk}^{l_{+}} \tau_{-},
\end{equation}
where $\tau_{ \pm} \equiv\left(\tau_{x} \pm i \tau_{y}\right) / 2$. $l = 1,2$ corresponds to the $p,d$ orbitals respectively. The hopping representations of these orbitals is given as
\begin{equation}
\lambda_{\bk}^{p_{+}}=\sum_{i=1}^{3} \w^{i-1} \sin k_{i}, \quad \lambda_{\bk}^{d_{+}}=\sum_{i=1}^{3} \w^{1-i} \cos k_{i}.
\end{equation}
Here $\w=e^{2 \pi i / 3}$ and $\lambda_{\bk}^{-}=\left(\lambda_{\bk}^{+}\right)^{*}$. The \(p\) and \(d\) orbitals obey the $E_{1u}$ and $E_{2g}$ representations respectively. \par

The underlying lattice Hamiltonians with lower angular momentum $l$ and therefore lower Chern numbers have more dominant band-inversion terms that suppress the onset of nematic ordering. This can be seen from a low-energy expansion that 
\begin{equation}
  (\Delta_\bk) = (k_x + i k_y)^l 
\end{equation}
\begin{figure*}
\centering
\includegraphics[width=1.0\textwidth]{ 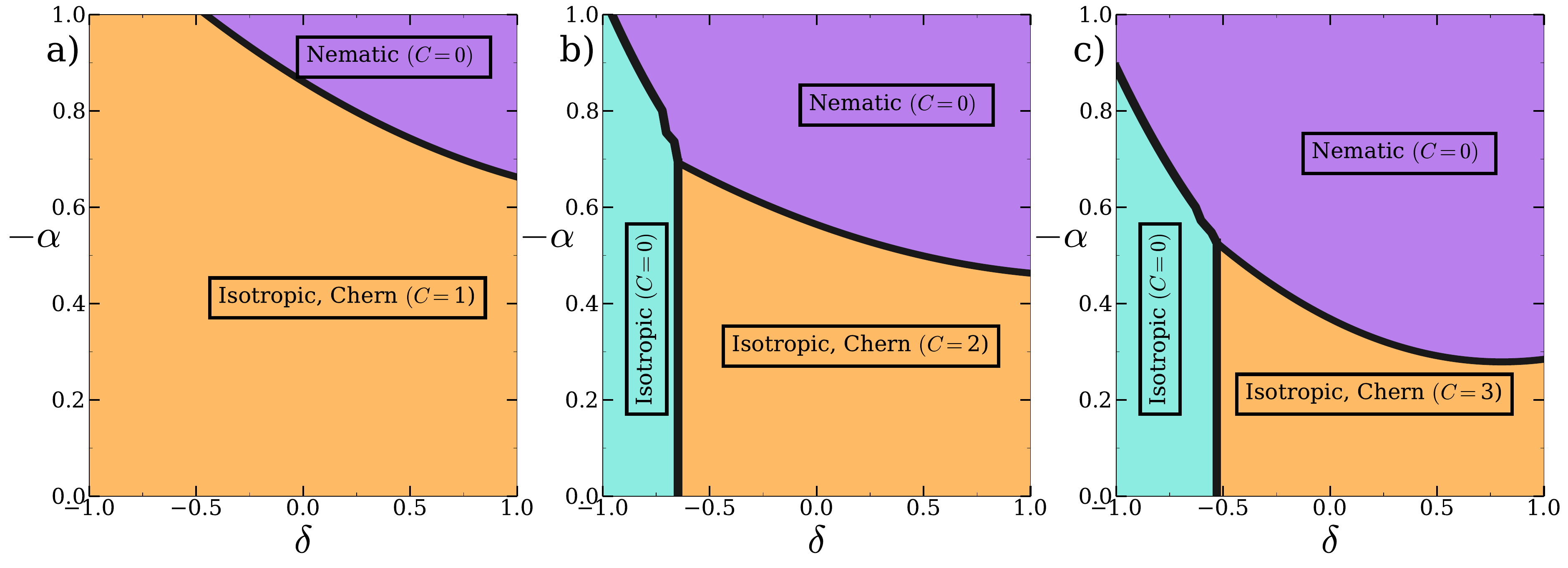}
\caption{The low-temperature phase diagram for $T/J = 1\times 10^{-3}$ and $\Delta = 1$ for a) $l = 1$, b) $l = 2$, and c) $l = 3$. The light blue region is an isotropic trivial state with $C = 0$.  Increasing the relative angular momentum of the orbitals leads to decreased relevance of the band-inversion and suppression of the Chern insulating state. }
\label{fig:angphase}
\end{figure*}
\end{document}